\documentstyle[pra,preprint,aps]{revtex}

\begin{document}
\draft
\title{Long Range Magnetic Order and the Darwin Lagrangian}
\author{Vishal Mehra and Jayme De Luca}
\address{Departamento de F\'{i}sica, Universidade Federal de S\~{a}o Carlos, 
\ \\
Rod. Washington Luiz km 235, 13565-905, Caixa Postal 676 , S\~{a}o Carlos,
SP,\\
Brazil}
\date{\today}
\maketitle

\begin{abstract}
We simulate a finite system of $N$ confined electrons with inclusion 
of the Darwin magnetic interaction in two- and three-dimensions. The lowest
energy states are located using the steepest descent quenching adapted for 
velocity dependent potentials. Below a critical density the ground state is 
a static Wigner lattice. 
For supercritical density the ground state has a non-zero kinetic energy.
The critical density decreases with $N$ for exponential confinement but not
for harmonic confinement. The lowest energy state also depends on the 
confinement and dimension: an antiferromagnetic cluster forms for
harmonic confinement in two dimensions.      
\end{abstract}


\section{INTRODUCTION}

The use of fast modern computers has made it increasingly easy to
investigate many-dimensional finite systems and
to study dynamical quantities like time to equipartition, symmetry break
in finite systems and quantities of thermodynamic interest
\cite{Chaos,unpubli,Cell,Thirring3,Thirring1,Michael}. The numerical studies 
have also revealed 
unexpected features of finite systems like cluster formation
\cite{Cell,Thirring3,Thirring1} and lack of equipartition\cite{Chaos,unpubli}.
 In this paper
we study numerically the energetic quenching of a classical electron gas
with inclusion of the magnetic interaction \cite{hanno,hannos,hannop}. 
The main motivation to study this system is the physics it describes and  
unfortunately the dynamics of this finite system is not easy to study 
because the equations of motion are algebraic-differential in 
character\cite{Petzold}. 
Because of this we explored numerically only the quenching motion, which 
involves integration of simple ordinary differential equations. 
Like the Coulomb interaction, the magnetic interaction is a long-range 
interaction. The long range nature of the Coulomb interaction has been widely 
explored in molecular dynamics and there is a large literature of numerical 
simulations and various special techniques were invented 
to deal with the long range nature of the interaction \cite{allen}. On the 
other hand, the magnetic interaction has been much less studied. 

 The magnetic interaction appears naturally in classical Maxwell 
electrodynamics: The lowest--order retardation and magnetic effects 
(or order $(v/c)^{2}$) can be described in terms of electron
variables only as a velocity--dependent interaction. This approximation was 
originally proposed by Darwin \cite{darwin} to obtain a lagrangian 
 which bears his name.The Darwin lagrangian is much--used
in atomic physics \cite{breit,landau} where it is known as 
Darwin--Breit interaction in its quantised form. The Darwin lagrangian
includes the lowest order correction to the electric field of a moving
charge and the lowest order magnetic field (the Biot-Savart
term). Apart from its traditional domain of atomic physics the Darwin
lagrangian has been used to model slightly relativistic plasmas \cite
{nielson,kaufman,ding} and even models of superconductivity and stellar 
magnetic 
fields \cite{hanno}. These ideas extend the range of Darwin lagrangian from
conventional relativistic correction: the Darwin corrections are also 
important in the low--energy (nonrelativistic) regime. One reason 
for that is to be found in the long--rangedness of the Darwin force 
and the fact that it is badly screened, unlike the Debye screening of 
the Coulomb force \cite{alastuey}. These two factors compensate for the 
weakness of the Darwin interaction at low velocities and make it a player 
in certain circumstances\cite{hanno,hannos}. The Darwin lagrangian has 
also been used as an unfolding of the scale invariant degenerate Coulombian 
interaction to estimate long-time-scale dynamical effects in atomic 
physics\cite{deluca}.  
Last, the Darwin lagrangian represents the 
first correction to the Coulomb interaction lagrangian in a series 
expansion of the Tetrode-Fokker lagrangian of the Wheeler-Feynman action at 
a distance theory. This relativistic lagrangian  theory has been shown to 
have interesting magnetic consequences\cite{Farrell} that deserve further
study.   

Experimentally, aggregates of electrons with the same sign of charge 
can be confined for long times using suitably chosen static electric 
and magnetic fields and form so--called nonneutral plasmas (For a recent 
review see \cite{neil_rmp}). Differently from neutral plasmas 
(i.e. composed of electrons with both
sign of charge in approximately equal numbers), the nonneutral plasmas can
attain thermal equilibrium and cooled to low enough temperatures to form
liquid and crystal-like states. These plasmas under various confinement
geometries have been extensively studied, particularly in the last decade.
Generally the interelectron interaction is taken to be Coulombic or Yukawa.

The study of the possible low--energy states of a system may shed some 
light on the possibility of a phase transition.
 We chose to study the magnetic long-range effects in a confined 
electron gas and in the neighborhood of its lowest energy
state. 
We confine a system of charged 
electrons in two and three dimensions by use of a
background field, taken either to be of a harmonic or exponential form. 
The electrons interact by Coulomb plus Darwin forces. For the purely 
Coulombian repulsion, it is known that the lowest energy state 
of this system is a Wigner lattice (triangular-like in two dimensions), 
which we also find 
with our quenching techniques \cite{peeters,ladir}.
 We show that the long range effects depend on $N$ and on a single 
parameter $\beta$, which is usually of the 
order of $10^{-1}-10^{-3}$ for attainable physical densities. 
We find that low-$N$ systems need an artificially 
larger value of $\beta$ for the long-range effects to be important.
The assumption is that the critical value goes down with $N$, and because 
of the practical impossibility of simulating systems with millions of
electrons, we investigate finite systems for an artificially higher 
value of $\beta$ and extrapolate the scaling properties to the 
large-$N$ case. This assumption holds for the exponential confinement but 
not for the harmonic confinement, where the critical $\beta$ does not go 
down with  $N$. 

This paper is divided as follows. The next section introduces the model and 
defines the quenching procedure. The  numerical results for the 2-D systems 
are described in the section~3; two subsections correspond to types of 
confining potential. The section~4 deals with the 3-D systems, again divided 
into subsections. The papers ends with a discussion in section~5. 

\section{Darwin lagrangian and the Natural Quenching}

We consider $N$ electrons in two or three dimensions 
interacting via the Coulomb repulsion plus the velocity-dependent 
Darwin magnetic interaction\cite{hannop,darwin} and
confined by a one-electron potential $V_{C}(\vec{r})$ of the positive
 background. The lagrangian for this system 
can be written as
\begin{equation}
L=\frac{m}{2}\sum _{i=1}^{N}\vec{v}_{i}^{2}-e^{2}\sum _{i<j}^{N}\frac{1}{%
r_{ij}} +e^{2}\sum _{i<j}^{N}\frac{\vec{v}_{i}.\vec{v}_{j}+(\vec{v}_{i}. 
\hat{e}_{ij})(\vec{v}_{j}.\hat{e}_{ij})}{2c^{2}r_{ij}}-\sum_i V_{C}(\vec{r}_i),
\label{lagrangian}
\end{equation}
where $\vec{r}_{i}$ and $\vec{v}_{i}$ are the position and the velocity of
the $i$th electron and $\vec{r}_{ij}\equiv\vec {r}_{i}-\vec{r}_{j}$, 
$r_{ij}\equiv|\vec{r}_{ij}|$,  $\hat{e}_{ij}\equiv\vec{r}_{ij}/r_{ij}$ is 
the unit vector pointing 
from the $i$th to the $j$th electron, $e$ is the  electronic charge, 
and $c$ is the velocity of light. The first term is the kinetic energy 
$E_{K}$ of the system and the second term is the Coulomb energy. The next 
term is the Darwin $V_{D}$ which can be expressed in terms of the vector 
potential $ \vec{A}$ as 
\begin{equation}
V_{D}=-\frac{e}{2c}\sum _{i}\vec{v}_{i}.\vec{A}_{i},
\end{equation}
with 
\begin{equation}
\vec{A_i}(r,v)\equiv e\sum _{j\neq i}\frac{\vec{v}_{j}+\hat{e}_{ij}( \vec{v}%
_{j}.\hat{e}_{ij})}{2cr_{ij}}.
\end{equation}
As the lagrangian is time-independent, there is an associated energy 
constant which evaluates to \cite{hanno} 
\begin{equation}
E=\sum_i\frac{1}{2}m\vec{v}_i^{2}+\frac{e}{2c}\vec{v}_{i}.\vec{A_i}%
(r,v)+\sum_{i<j}{{e^2}\over{r_{ij}}} +\sum_i V_C(\vec{r}_i) .
\end{equation}
Notice that this energy is $\emph{not}$ of the minimal coupling type
\begin{equation} 
E=\sum {{1}\over{2}}(p_i-A_i)^2 +V,
\label{minimal}
\end{equation}  
which is only the case when the magnetic field is external to the system,
and consequently the vector potential is velocity independent. In the present 
situation, because of the internal fields, the state of minimal energy is not 
always the zero velocity case anymore, as the conditions for the 
Bohr-Van Laufen theorem are not satisfied \cite{hannop,alastuey,vleck}
(This theorem states that a velocity independent vector potential does not 
affect the partition function $\emph{if}$ the energy is of the minimal 
coupling form (\ref{minimal})).

 The form of the equations can be simplified by using
 scaled units:  a length scale, 
given by the average interelectron separation $R$, scales positions as 
$\vec{x}\rightarrow R\vec{x}$ (in this units the
gas has a density equal to one). Time is scaled as 
$dt\rightarrow\omega_{0}d\tau$ where $\omega_{0}^{2}\equiv e^{2}/m R^{3}$. 
In these units the energy scales as $E\rightarrow m(\omega_0 R)^{2}\hat{E}$ 
with 
\begin{equation}
\hat{E}=\sum_i\frac{1}{2}\vec{v}_i^{2}+\sum_{i<j}{{1}\over{r_{ij}}}
+\beta^{2}\sum_{i<j} \frac{\vec{v}_{i}.%
\vec{v}_{j}+(\vec{v}_{i}.\hat{e}_{ij}) (\vec{v}_{j}.\hat{e}_{ij})}{2r_{ij}}
+\sum_i \hat{V}_C (\vec r_i) .
\label{energy}
\end{equation}
The parameter $\beta^{2}$ in the above equation is defined as 
\begin{equation}
\beta^{2}\equiv {{r_e}\over{R}},
\end{equation}
where $r_e =e^2/mc^2$ is the classical electronic radius and the 
interelectron distance $R$ in 2-D is given by $R=1/\sqrt{n}$ and
$R=n^{-1/3}$ in 3-D.
For some real physical situations: The conduction band in metals forms 
a 3-D degenerate plasma with typical densities of $n \sim 10^{23}$ cm$^{-3}$,
which gives for $\beta^2$ the value of $\beta^2 \sim 10^{-6}$. The highest 
density physical plasma is found in the interior of white dwarf stars, 
corresponding to an electron density of $n \sim 10^{32}$ cm$^{-3}$ which
gives $\beta^2 \sim 10^{-3}$ \cite{Ichimaru}.
   
Once the systems studied here are rotationally invariant, 
Noether's theorem determines a constant of motion\cite{Arnold} 
for them 
\begin{equation}
C\equiv \sum _{i=1}^{N} \vec{r_i}\times \vec{v}_{i} \
+{\beta}^{2}\sum _{i<j}^{N}\frac{\vec{r}_{i}\times \vec{v}_{j}
+(\vec{r}_{i}\times 
\hat{e}_{ij})(\vec{v}_{j}.\hat{e}_{ij})}{2r_{ij}},
\label{noether}
\end{equation}
with $\hat{e}_{ij}\equiv \vec{r}_{ij}/r_{ij}$, as before. For 2-D this 
constant is a vector perpendicular to the plane and for 3-D the $O(3)$
symmetry determines a constant vector by the above formula. This constant 
can be interpreted simply as the sum of the mechanical angular momenta 
plus the field angular momentum.

\par
To look for the minimum energy states of the velocity dependent $N$-body 
system, we adapt a numerical procedure analogous to the steepest descent 
quenching using what we name the natural quenching vector field. 
We check that for potential systems this procedure produces the static 
crystalline arrangement of electrons known as the Wigner lattice 
\cite{peeters,ladir}.
We now define the natural quenching vector field, which is constructed from 
the differential of the expression for the energy constant given
by equation (\ref{energy}).
We start from a random initial condition and integrate it as a function
of a ``quenching parameter'' by the following gradient equations:
\begin{eqnarray}
{{d\vec {r}_i}\over{ds}} = - {{\partial E}\over{\partial \vec{r}_i} } , 
\;\; {{d\vec {v}_i}\over{ds}} = - {{\partial E}\over{\partial \vec{v}_i} } .
\label{quenching}
\end{eqnarray}
It is easy to see that along this gradient motion the energy always decreases,
 as the parameter derivative of the energy evaluates to
\begin{eqnarray}
{{dE}\over{ds}}=-\sum_i |{{\partial E}\over{\partial \vec{r}_i} }|^2 
-\sum_i |{{\partial E}\over{\partial \vec{v}_i} }|^2.
\end{eqnarray}

Along with the numerical quenching from random initial conditions 
it is necessary to use the relativistic form of the kinetic energy. 
Otherwise, we observe that some electrons acquire an enormous kinetic 
energy during quenching,       
creating an enormous nonphysical internal field that still decreases the total
energy. Of course, for such large velocities the Darwin approximation breaks 
down and the whole lagrangian describes nonphysical effects, as 
discussed in reference \cite{Bessonov}. In all our numerical experiments we 
check that the electron energies were never relativistic in the final 
quenched state, which guarantees that the Darwin approximation is valid.
Last, to gain some understanding of how the above quenching 
procedure can find states with nonzero velocity, let us examine
equation (\ref{quenching}) for the velocities, which read as
\begin{eqnarray}
\label{quenchmatrix}
{{d\vec {v}_i}\over{ds}} = -\vec{v}_i -\beta^2 \sum_{j \neq i} 
\frac{\vec{v}_{j}+ \hat{e}_{ij}(\vec{v}_{j}.\hat{e}_{ij})}{2r_{ij}}.
\end{eqnarray}
Notice that on the right hand side we have a linear function
of the velocities, defining a linear matrix $M(\vec{r}_i,\beta^2)$.
For $\beta^2=0$, this matrix is minus the identity and the velocities
are all quenched down to zero. Above a critical value of $\beta^2$, this
matrix can have negative and zero eigenvalues, and it is not possible
to quench the velocities to zero anymore, which is the cause of the
nonzero velocity states we find. 
The critical point $\beta^{2}_c$ can also be located by an alternative 
analytical method: Consider equation (\ref{quenchmatrix}) for the 
velocity-quenching.
 For $\beta^{2}=0$, the eigenvalues of $M$ are all degenerate and equal 
to one. Taking the electron coordinates to be those of the static Wigner 
lattice (which can be obtained for $\beta^{2}=0$), one can diagonalize $M$ 
numerically and find all its eigenvalues. The critical $\beta^{2}_c$ is that 
for which the minimum eigenvalue of $M$ crosses zero i.e. the minimum 
eigenvalue of $M$ is just negative. It can be  seen then that in this case 
the quenching will decrease the energy $\emph{while}$ increasing the 
velocities along the negative eigenvector directions.  We have diagonalized 
$M$ in the neighborhood of the Wigner lattice and it is satisfactory that the 
critical $\beta^{2}$ calculated by the matrix method agrees with the values 
obtained by quenching.

\section{Numerical Results for Circular disk geometry}

\subsection{Harmonic confinement}
\label{harmonic}
 We consider first  a system of   $N$ electrons in 2-D, confined 
by the field of a uniformly charged circular disk of positive charges, 
of radius $R_d$ scaled units, and with the electronic density
of one electron per squared scaled unit ($N=\pi R_d^2$). For this
system the potential of the uniformly charged disk of positive background 
can be calculated analytically\cite{morse} and for $r<R_d$ it is  
approximated by 
\begin{equation}
\label{har_confine}
\hat{V}_C(r)=-2\sqrt(\pi N)+{{\pi^{3/2}}\over{2N^{1/2}}}r^2.	
\end{equation} 
We have explicitly included the negative constant to properly account for 
the electrostatic interaction with the positive background. 
One still needs to add  the self--energy of the positive background 
$\sqrt{\pi}N^{3/2}/8$ to expression (\ref{energy}) in order to get the total 
electrostatic energy. 
We seek to determine the lowest--energy states of this system by employing  
the natural quenching technique described above, and we integrate equations
 (\ref{quenching}) numerically with an 6/7 Runge-Kutta embedded integrator 
pair. By quenching from different 
initial conditions we can hope to obtain insight into the character of the
ground state.  The natural quenching is performed for the disk system for 
various
values of the parameter $\beta^{2}$. The electrons are
started  from a triangular lattice, distorted slightly in a random manner, 
with velocities uniformly (and randomly) distributed upto a certain maximum 
value. A square lattice type initial configuration is also used and the same 
final result is found. The system is quenched until a steady state appears to 
have been reached. To check if we actually attain a global minimum state and 
not merely a local minimum, we slightly heat the obtained configuration and 
quench it again. By these means we are confident that our ground states are 
at least qualitatively correct. 

These simulations are done for 225 to 1600 electrons in the disk. In all 
cases it is observed that below a certain value of the parameter $\beta^{2}$ 
the ground state is the static Wigner lattice, independent of the $\beta^{2}$.
 But above the critical $\beta^{2}_c$, a new type of ground state is obtained.
 This state has nonzero kinetic energy with a striking nonuniform distribution
 of velocities. The electrons with large velocities are confined to an 
antiferromagnetic cluster in the center of the disk. The configuration in 
the position space remains visibly triangular-like. The electrons in the 
central cluster have velocities aligned in a manner to minimize the Darwin 
energy (Fig.\ref{P900}). This parallel and antiparallel orientation of the 
velocities succeeds in lowering the energy of the nonstatic configuration 
below the static Wigner lattice. 

The critical parameter value $\beta^{2}_c$ decreases with the increasing  $N$,
 the number of electrons in the disk, but it is rather a weak dependence. 
For the 225 electrons disk, $\beta^{2}_c\approx 0.72$ while it is 
approximately  0.71 for the 1600 electrons disk. The relative size of the 
cluster slowly decreases with increasing $N$: we quantify it as following.  
A electron $i$  {\it belongs} to the cluster if $\beta^{2}v^{2}_i>0.01$. The 
quantity $\beta^{2}v^{2}$ is generally about 0.06 for the fastest electron. 
At $\beta^{2}=0.75$, this  criterion yields the fraction of the cluster 
electrons to be 0.31 for the 225 electron system and monotonously decreasing 
to 0.26 for $N=900$ and to 0.23 for $N=1600$. However it appears that the 
cluster remains equally hot, independent of $N$, i.e., the average kinetic 
energy per cluster electron does not depend on $N$ at constant $\beta^{2}$. 
Futher increasing the value of $\beta^{2}$ beyond $\beta_c^2$ causes rapid 
increase both in the 
size and the temperature of the cluster. The ground--state energy continues 
to decrease as $\beta^{2}$ is increased beyond $\beta^{2}_c$ 
(Fig.~\ref{figenergy}).

The quenching runs do not always yield the ground state. Often, in particular 
for larger sized disks, an imperfectly aligned higher energy state is 
obtained. The local order is of the same type as the true ground state but 
on a larger scale two or more regions of the local order have an mismatch, 
analogous to grain--boundaries in a polycrystalline material 
(Fig.~\ref{P900B}).

\subsection{Exponential confinement}
The effect of choice of the confining potential may be studied by 
considering a different potential. To this end, we replace the harmonic 
potential by an exponential potential,
$$V_C=V_0\exp((r-R_d)/r_W).$$
Here, as before $R_d=\sqrt(N/\pi)$ and $r_W=0.5$ in scaled units. The 
quenchings are performed in a manner identical to that described above, 
starting from a randomly distorted triangular lattice. The critical 
$\beta^{2}_c$ is obtained, separating the static and 
non--static ground--states. The values of $\beta^{2}_c$ are rather lower and 
they {\it decrease}  appreciably with increasing $N$. We find 
$\beta^{2}_c\approx 0.52$ for $N=225$ to about 0.42 for $N=1600$.  
These (and the intermediate $N=484$ and $N=900$) values may be fitted to 
a power-law 
$\beta^{2}_c\sim N^{\alpha}$ with $\alpha\approx -0.11$. 

The character of the ground--states is also affected. The edge (i.e. the 
surface) of the static lattice is no longer triangular--like but is composed 
of two ring--like layers. Above $\beta^{2}_c$ the velocity distribution is 
highly inhomogeneous: the kinetic energy is concentrated in the two edge 
layers (Fig.~\ref{900F}). Increasing $\beta^{2}$ beyond $\beta^{2}_c$ does 
not result in more electrons acquiring kinetic energy but merely increases 
kinetic energy of the edge electrons.

\section{Numerical Results for Spherical Geometry}
\subsection{Harmonic confinement}

In this section we consider $N$ electrons confined by the field of a 
homogeneous positively charged sphere of radius $R_d$. The potential 
(for $r<R_d$) can be calculated exactly,
$$V_C=-2\pi R_d^{2}+\frac{2\pi}{3}r^{2},$$
which follows immediately from Gauss law of electrostatics. To obtain the 
total electrostatic energy we must again add the self--energy contribution 
of the background $\frac{\pi}{5}NR_d^{2}$ to the expression (\ref{energy}).
As before, the electron density is taken to be one electron per scaled unit. 
Quenchings are performed for $N$=216-1000 electron systems. Simulations are 
started from a randomly distorted cubic lattice, while the velocities are 
initialized in the manner previously described for the disk geometry. As in 
the disk geometry, a critical $\beta^{2}$ separates static ground states 
from the nonstatic ground states. The critical $\beta^{2}_c$ is slightly 
smaller in three dimensions--- it varies from $\approx 0.67$ for $N=216$ 
to about 0.66 for $N=1000$. 

But the character of the ground states is very different from that obtained 
for the disk geometry. The electrons are arranged in a multiple ring--like 
structure around the center of the ball. These rings possess sharp boundaries 
and their number grows with $N$. For example, the 216 electrons system has 
4 such rings (including the cluster of central electrons) while 7 rings are 
visible for $N=1000$ system.  

This ring--like structure persists beyond $\beta^{2}_c$. Though, the velocity 
distribution is not homogenous, a distinct cluster of hot electrons is not 
formed.  The velocities project radially outwards (Fig.~\ref{3d216}). The 
electrons in the small central cluster have  smaller than average kinetic 
energy. The kinetic energy is fairly shared between the other rings but 
the distribution between electrons in a particular ring is non--uniform.

The different ordering in two and three dimensions has striking effects: 
in 3-D the Darwin interactions between neighboring electrons are highly 
repulsive and the lowering of the energy is provided by the Darwin 
interactions of distant electrons. Whereas, in the 2-D disk geometry, 
the Darwin interactions between nearest neighbor electrons lower the 
Darwin energy.

\subsection{Exponential confinement}
In this subsection we describe the results of the simulation of 3-D 
electron gas confined by an exponential potential
 $$V_C=V_0\exp((r-R_d)/r_W).$$
Here $R_d$ is defined by the relation $N=\frac{4\pi}{3}R_d^{3}$, and 
$r_W=1.5$. The static and nonstatic ground states are again obtained, 
below and above $\beta^{2}_c$ respectively. Though, in contrast with the 
harmonic confinement, the $\beta^{2}_c$ decreases with increasing $N$ with 
the power--law $\beta^{2}_c\sim N^{-0.23}$. The values of $\beta^{2}_c$ 
range from 0.74 for $N=216$ to 0.50 for $N=1000$. 
            
The lattice has a ring--like structure again but with some differences. 
First, the number of rings is smaller: $N=1000$ system has only 4 rings 
compared with 7 with harmonic potential. Second, the central cluster is 
absent. The distribution of kinetic energy is also different from the 
harmonic case. The kinetic energy per electron increases as one goes 
outwards and for 
$N=1000$ the inner two rings have  almost no kinetic energy.

\section{Discussions and Conclusions}

It would be natural to integrate the supercritical 
symmetry-breaking states of low-energy and nonzero angular momentum 
to see the time dependence of the angular momentum \cite{Michael},     
but this is not an easy numerical job.
 The existence of zero and negative eigenvalues of $M$ in (\ref{quenchmatrix}) 
signal the onset of complex dynamical behavior for this system: 
The lagrangian equations of motion that follow from (\ref{lagrangian}) are
\begin{eqnarray}
\vec{a}_i +\beta^2 \sum_{j \neq i} 
\frac{\vec{a}_{j}+ \hat{e}_{ij}(\vec{a}_{j}.\hat{e}_{ij})}{2r_{ij}} & 
= & -{{dV_C}\over{d\vec{r}_i}}-\sum_{j\neq i}{{\hat{e}_{ij}}\over{r_{ij}}} 
\nonumber \\
+\beta^2 \sum_{j\neq i}{{1}\over{2r_{ij}^2}}[ (\vec{v}_i\cdot \hat{e}_{ij})\vec{v}_j+(|\vec{v}_j|^2-3\vec{v}_j\cdot\hat{e}_{ij} -2\vec{v}_i \cdot \vec{v}_j)\hat{e}_{ij} ].
\end{eqnarray}
Notice that this is an algebraic-differential equation of order zero
\cite{Petzold}, and the linear matrix sitting on the left side is the
same exact matrix $M$ that appeared in (\ref{quenchmatrix}). The numerical 
procedure to integrate this equation is delicate: If the 
matrix has maximal rank, which is the case for low values
of $\beta$, the integrator RADAU\cite{Hairer} can be used. If the matrix
has only one zero eigenvalue, then the integrator DASSL\cite{Petzold} can 
be used, as long as the matrix does not lose rank along the trajectory, 
which is the case for rare initial conditions only. The general case 
where the matrix's rank is lesser than $2N-1$, or even worse if it loses 
rank along the trajectory, then one is faced with a rich 
system which could generate a very complex dynamics. As a matter 
of fact, in 1976, one of the earliest studies of this many-body system 
declared it intractable\cite{nielson} and a coarse grained field 
approximation was developed to study it, which later became a plasma 
simulation technique \cite{ding}. We could integrate initial conditions
with a very low $\beta$, and there we found that the angular momentum is an 
approximate constant, as it should be for the $\beta=0$ case, according to
equation (\ref{noether}). For 
intermediate values of $\beta$ we were able to perform the numerical 
integration of the dynamics using DASSL, and we find that the total angular 
momentum does not change sign for very long time scales. In the supercritical 
situation, where it would be of interest to study the dynamics, we find that 
the time steps of RADAU quickly go to zero due to the criticality of the 
matrix. An integration method has still to be developed to simulate
this interesting dynamics.   

In three dimensions, because of Gauss law, charge neutrality causes the 
total electrostatic energy to be an extensive quantity. As a matter of fact, 
our data show that $E/N$ tends to a constant value in 3-D. On the other hand  
our data for 2-D indicates that the total electrostatic energy is 
non--extensive.

 At constant $\beta^2$ the Darwin lowering of energy is smaller for larger 
$N$ systems in 2-D. This can also be inferred from the observation made in 
section~3A that the fraction of hot electrons goes down with increasing $N$. 
Thus our results indicate that as $\lim N\rightarrow\infty$ at constant 
electron density the Darwin lowering vanishes. This conclusion holds for 
harmonically confined systems for which $\beta^{2}_c$ is almost independent 
of $N$ and hence different sized systems can be compared at constant 
$\beta^2$. 

The magnetic lowering does not decrease with large $N$ in 3-D. This makes the 
question of a proper thermodynamic limit  more subtle. In a certain sense, 
the Darwin interaction merely renormalizes the electronic charge. But this 
would interfere with the cancellation of the background charge and hence 
jeopardize the thermodynamic limit \cite{ruelle}.  We feel that more 
numerical and analytical work is needed to resolve the question of the 
thermodynamic limit for the Darwin lagrangian.

Acknowledgements: We acknowledge discussions with A. Castelo, F. C. Alcaraz 
and J. P. Rino.

\begin{figure}

\caption{The ground-state of harmonically confined disk with $N=900$ at 
$\beta^{2}=0.75$. The electrons are located at the tails of the arrows; 
the lenght of the arrows is proportional to the magnitude of the electron 
velocities obtained after quenching. The direction of an arrow gives the 
angle of the corresponding velocity vector. Scaled coordinates.}
\label{P900}    
\end{figure}

\begin{figure}  

\caption{(a) Ground--state energy of harmonically confined disk with 
$N=225$ vs. $\beta^{2}$. The critical $\beta^{2}$ is slightly less than 0.72.
(b) Ground--state energy of harmonically confined electron gas in 3-D with 
$N=216$ vs. $\beta^{2}$. The critical $\beta^{2}$ is slightly less than 0.67.}

\label{figenergy}
\end{figure}

\begin{figure}

\caption {A higher-energy minimum of harmonically confined disk with 
$N=900$ at $\beta^{2}=0.75$. The arrows are made in the manner described 
in the previous figure.}
\label{P900B}

\end{figure}

\begin{figure} 

\caption{ The ground--state of exponentially confined electron gas in 2-D 
at $\beta^{2}=0.45$. The arrows are drawn as in Fig.~1. }

\label{900F} 

\end{figure}

\begin{figure}  

\caption{The ground-state for harmonically confined system in three 
dimensions with $N=216$ at $\beta^{2}=0.74$. The velocities project 
radially outwards. Arrows are made in the similar manner to the 
two-dimensional case.}
\label{3d216}

\end{figure}

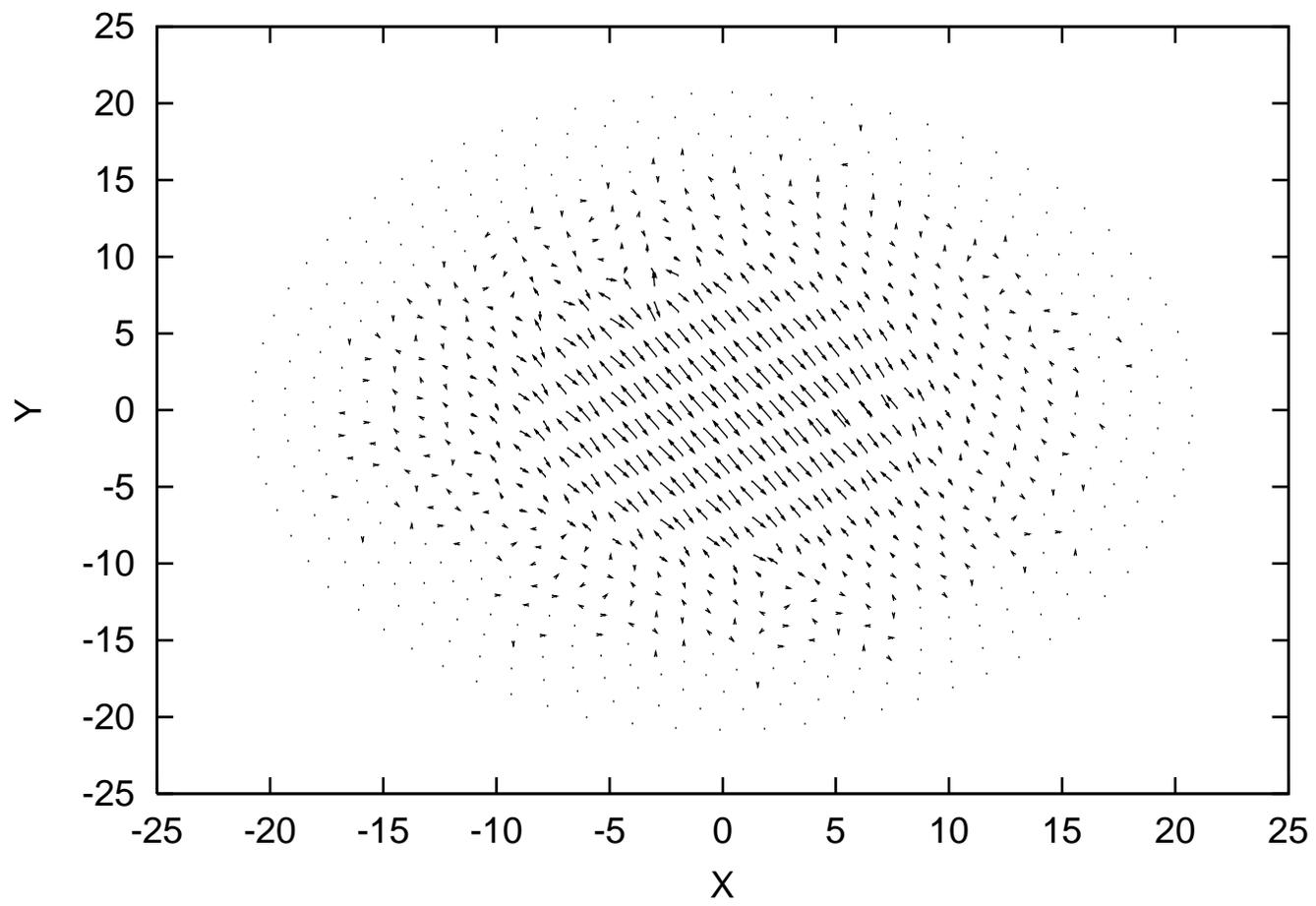
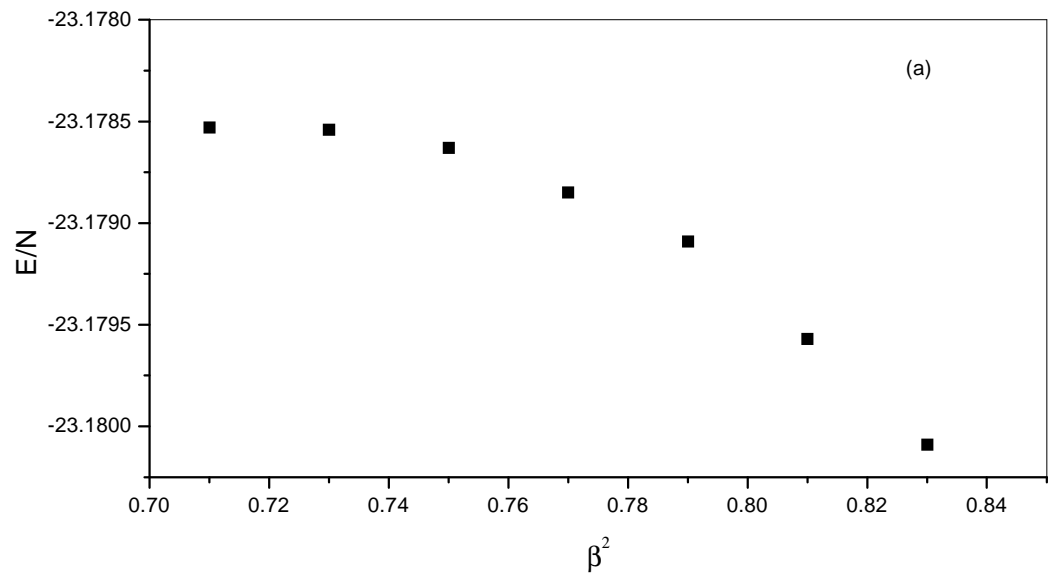
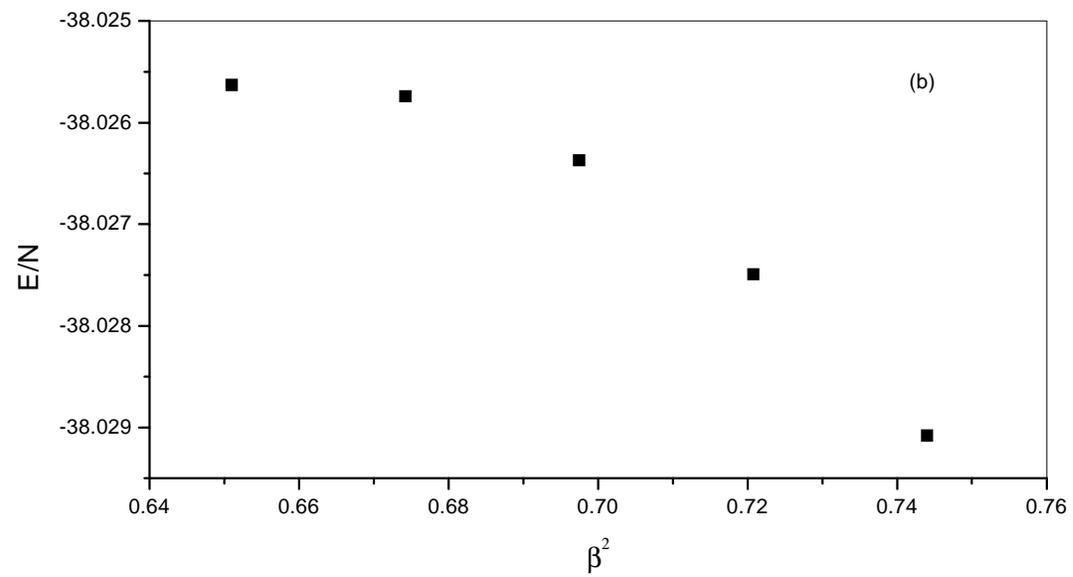
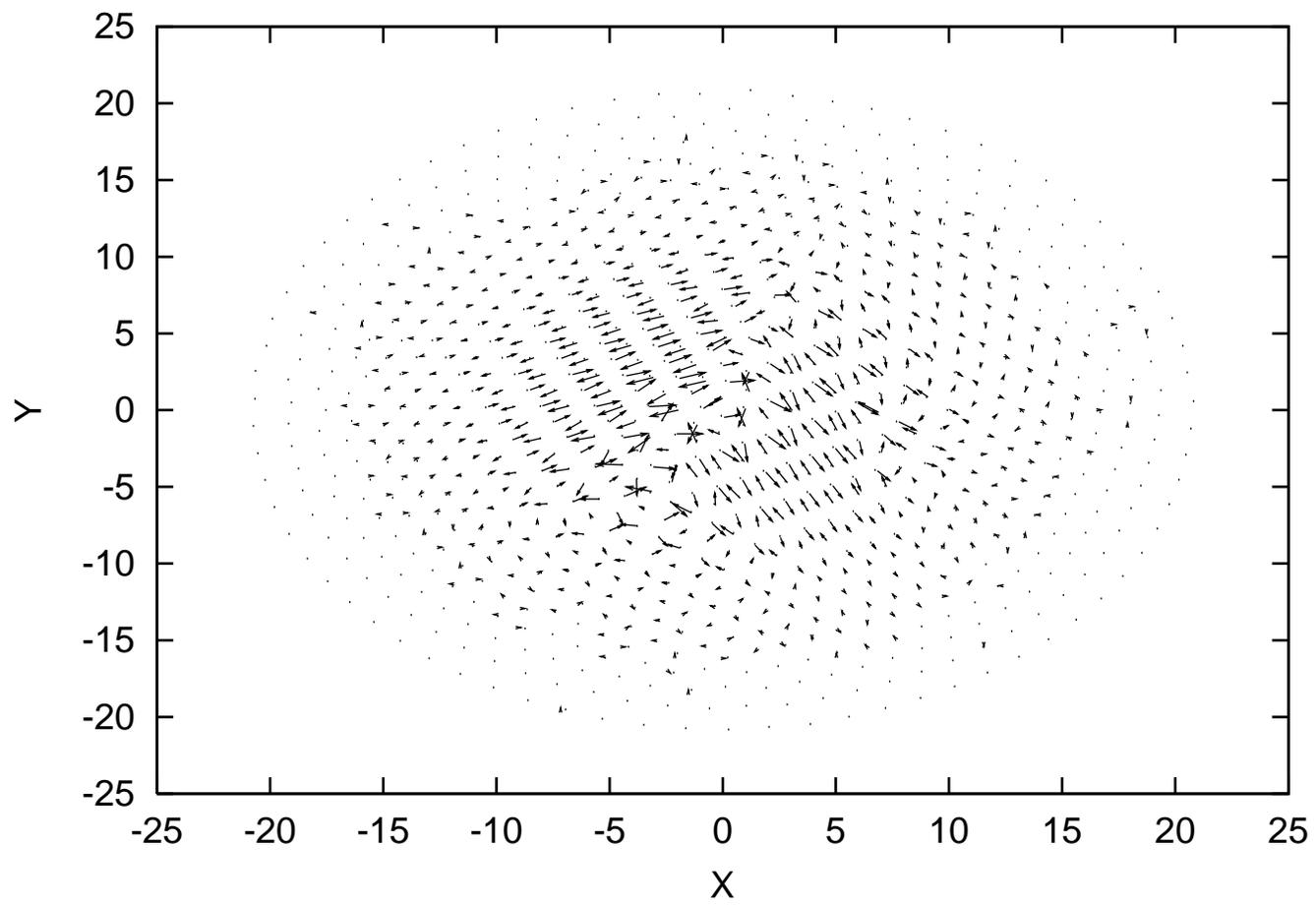
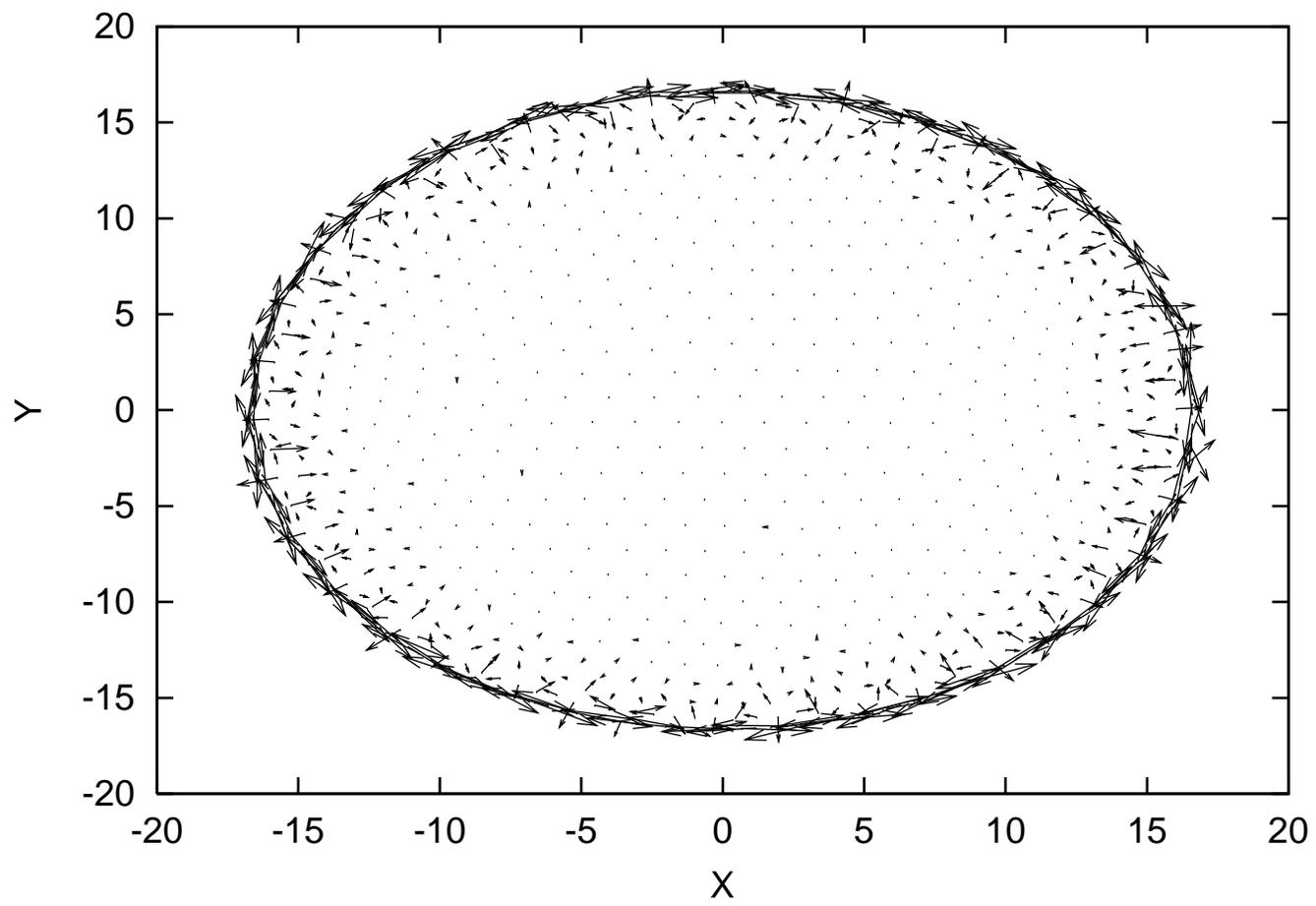
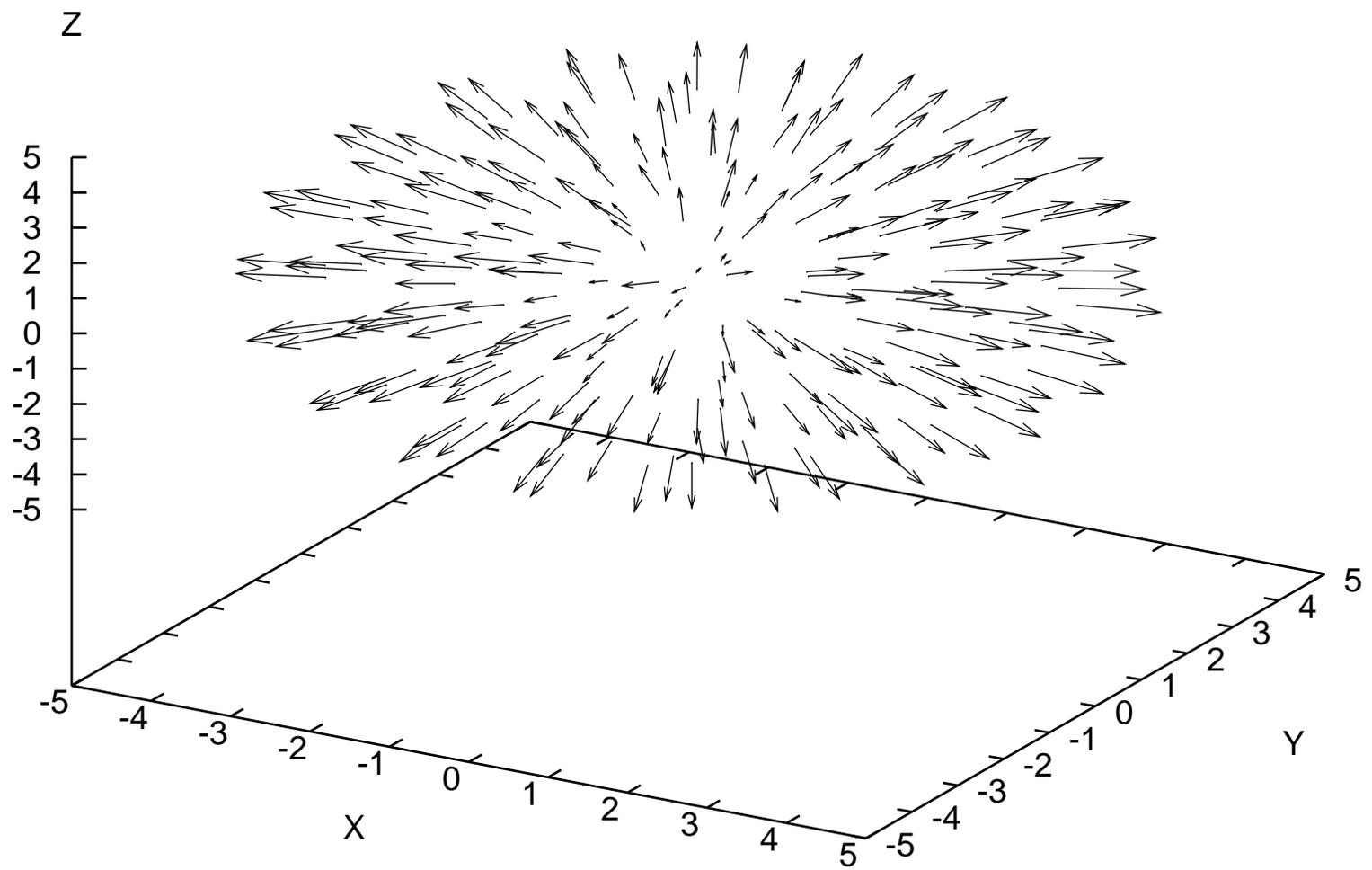
\end{document}